\documentclass[aps,prc,reprint,superscriptaddress]{revtex4-1}

\usepackage {graphicx}
\usepackage{units}
\usepackage{amssymb}
\usepackage{amsmath}
\usepackage{hyperref}
\usepackage{bbm}

\def\D{\mathrm{d}}

\newcommand{\tsc}[1]{\textsc{#1}}
\newcommand{\Py}{\tsc{Pythia}}

\begin{document}




\title{Strangeness Enhancement due to String Fluctuations}


\author{H.J.~Pirner}
\affiliation{Institute for Theoretical Physics, Heidelberg University, Germany}

\author{B.Z.~Kopeliovich}
\affiliation{Departemento de Fisica, Universidad Tecnica Federico Santa Maria, Chile}

\author{K.~Reygers}
\affiliation{Physikalisches Institut, Heidelberg University, Germany}



\begin{abstract}
We study string fragmentation in high multiplicity proton-proton collisions in a model where the string tension fluctuates. These fluctuations produce exponential pion spectra which are fitted to the transverse momentum distributions of charged particles for different multiplicities. For each multiplicity the so obtained hadronic slope parameter defines the magnitude of the string fluctuations which in turn determines the produced ratio of strange to light quarks. \Py\ string decay simulations are used to convert each ratio of strange to light quarks to the appropriate ratio of strange hadrons to pions.
\end{abstract}



\maketitle



\section{Introduction}
\label{sec:intro}

Recently hadronic spectra with strange and multistrange hadrons were measured in pp collisions at the LHC \cite{ALICE:2017jyt}. With increasing multiplicity a strong enhancement of strangeness was observed. This result suggests collective processes in pp collisions which were advocated for heavy ion collisions in Refs. \cite  {Koch:1986ud, Koch:2017pda}. In this paper we want to present an alternative approach  focusing on the string dynamics in low momentum hadron production. Our work parallels other recent work on modifications of string dynamics \cite {Fischer:2016zzs,Bierlich:2014xba,Bierlich:2015rha}.

Previously we have shown that in AA collisions where many strings are produced, flux tube dynamics can influence the azimuthal symmetry of the produced hadrons \cite {Pirner:2014zka}. In this work we did not relate the Gaussian momentum distribution of quarks obtained from the Schwinger model \cite {Schwinger:1951nm} and the tunneling mechanism of Casher, Neuberger, and Nussinov \cite{Casher:1978wy} with the observed exponential hadron spectra. This difference has been discussed in the literature \cite{Bialas:1999zg} and the idea is that  fluctuations of the string tension (the energy per unit length of the tube) can account for the ``thermal'' distribution of hadrons. The spectrum of primarily produced hadrons would then be close to the maximum entropy distribution \cite {Pirner:2011ab}. There would be no need for further collisions between partons to obtain the final form of the observed distribution. Such a mechanism could explain early thermalization. Obviously, pp collisions present good examples to test this hypothesis, since in these collisions the available interaction volume and interaction time is limited.

In this paper we will work out the details of this idea in three stages. We first calculate the effect of string tension fluctuations on the Gaussian transverse momentum spectrum of produced quarks in the Schwinger model. Then we take into account that the mean transverse momentum of the produced hadron arises from the transverse momenta of the produced quark and antiquark, i.e., it is larger than the transverse momentum of the quarks. As a second step, we fit this theoretical form of the spectrum to the observed hadron spectra. The data indicate that the fluctuations of the string tension become larger with increasing hadron multiplicity. In the third step we use the ratios of produced strange to light quarks as input to a \Py\ calculation of string fragmentation. This way each strange and multistrange hadron is calculated from the string fragmentation as implemented in the standard \Py~8.2 code. The fluctuations with higher average string tension naturally produce relatively more strange quarks. The result of the third stage is then calculated as a function of the total charged-hadron multiplicity and compared with data.

\section{Hadronic fragmentation spectrum from a fluctuating flux tube}
Flux tubes decay by producing quarks and antiquarks tunneling from the vacuum. The resulting transverse spectrum of quarks has the form \cite{Casher:1978wy}
\begin{equation}
\frac {\D n}{\D^2p_\mathrm{q,\perp}} \propto e^{-\frac{\pi m_\mathrm{q,\perp}^2}{\kappa}},
\label{eq:schwinger_formula}
\end{equation}
where $\kappa$ is the string tension, the energy stored in the tube per unit length, and $m_\mathrm{q,\perp}=\sqrt{m_\mathrm{q}^2+p_\mathrm{q,\perp}^2}$ is the transverse mass of the quark. Conventionally, the quark masses which enter this formula are the constituent masses, i.e., for the light u and d quarks we use the constituent mass 
\begin{equation}
  m_\mathrm{q} = \unit[0.3]{GeV}.
  \label{eq:quark_masses}
\end{equation}
This quark mass is approximate, and we did not try to optimize its value for a best fit to the data. The transverse flux tube size varies as $r^2 \propto 1/\kappa$. Defining $\lambda^2 \equiv \kappa$ and following the suggestion of Bialas \cite{Bialas:1999zg} we let $\lambda$ fluctuate according to a Gaussian distribution. More general weight functions are possible. In the context of Tsallis distributions it has been suggested that gamma distributions are appropriate as weight functions \cite{bcr}. Here we proceed with Gaussian fluctuations:
\begin{equation}
P(\lambda)\,\D\lambda = \sqrt{\frac{2}{\pi \mu}}e^{-\frac{\lambda^2}{2 \mu}} \, \D \lambda.
\label{eq:gaus_fluct}
\end{equation}
The average string tension is 
\begin{equation}
\langle \kappa \rangle \equiv \langle \lambda^2 \rangle = \mu = \int_0^{\infty} \lambda^2 P(\lambda) \, \D\lambda.
\end{equation}

Taking into account the Gaussian fluctuations as described by Eq.~\ref{eq:gaus_fluct}, one obtains an exponential transverse momentum distribution for the produced quark or antiquark:
\begin{equation}
\frac{\D n}{\D^2p_{\mathrm{q},\perp}} \propto  e^{- \sqrt{\frac{2\pi (m_\mathrm{q}^2+p_{\mathrm{q},\perp}^2)}{\langle \kappa \rangle}}}.
\label{eq:quark_spectrum}
\end{equation}
A meson combines a quark and antiquark originating from two different strings breaks (mostly successive). Assuming, as usual, that those breaks are independent and controlled by the Schwinger mechanism (including fluctuations), we arrive at additive mean squared values of transverse momenta:
\begin{equation}
\langle p_\perp^2 \rangle = \langle p_{\mathrm q, \perp}^2 \rangle + \langle p_{\bar{\mathrm{q}},\perp}^2 \rangle.
\end{equation} 
This leads to the following spectrum for pions:
\begin{equation}
\frac {\D N}{\D^2p_\perp} = N_0 e^{- \sqrt{\frac{2\pi (m_\mathrm{q}^2+p_\perp^2/2)}{\langle \kappa \rangle }}}.
\label{eq:meson_spectrum}
\end{equation}

The mean transverse momentum squared of the meson of this distribution equals the sum of the mean transverse momenta squared of the quark and antiquark. The spectrum does obey $m_{\bot}$ scaling, but with a meson mass which is 1.5 times higher than the constituent quark mass. This spectrum does not  take into  account chiral symmetry associated with the small mass of the pion and therefore leads to a worse quality fit at small momenta. But we think that the slope of the spectrum around $p_\perp \approx 1~\mathrm{GeV}/c$ is a good indicator of the quark fragmentation dynamics underlying it. In principle large differences of quark momenta are influenced by the wave function of the meson which is not taken into account in the simple formula above. A full calculation including the folding of the quark distributions with the wave function of the meson is worked out in the appendix. It gives a result very similar to the simplified procedure. 

\section{Fitting hadronic spectra for high multiplicities in pp collisions}
\begin{figure}
\centering
\includegraphics[width=0.95\linewidth]{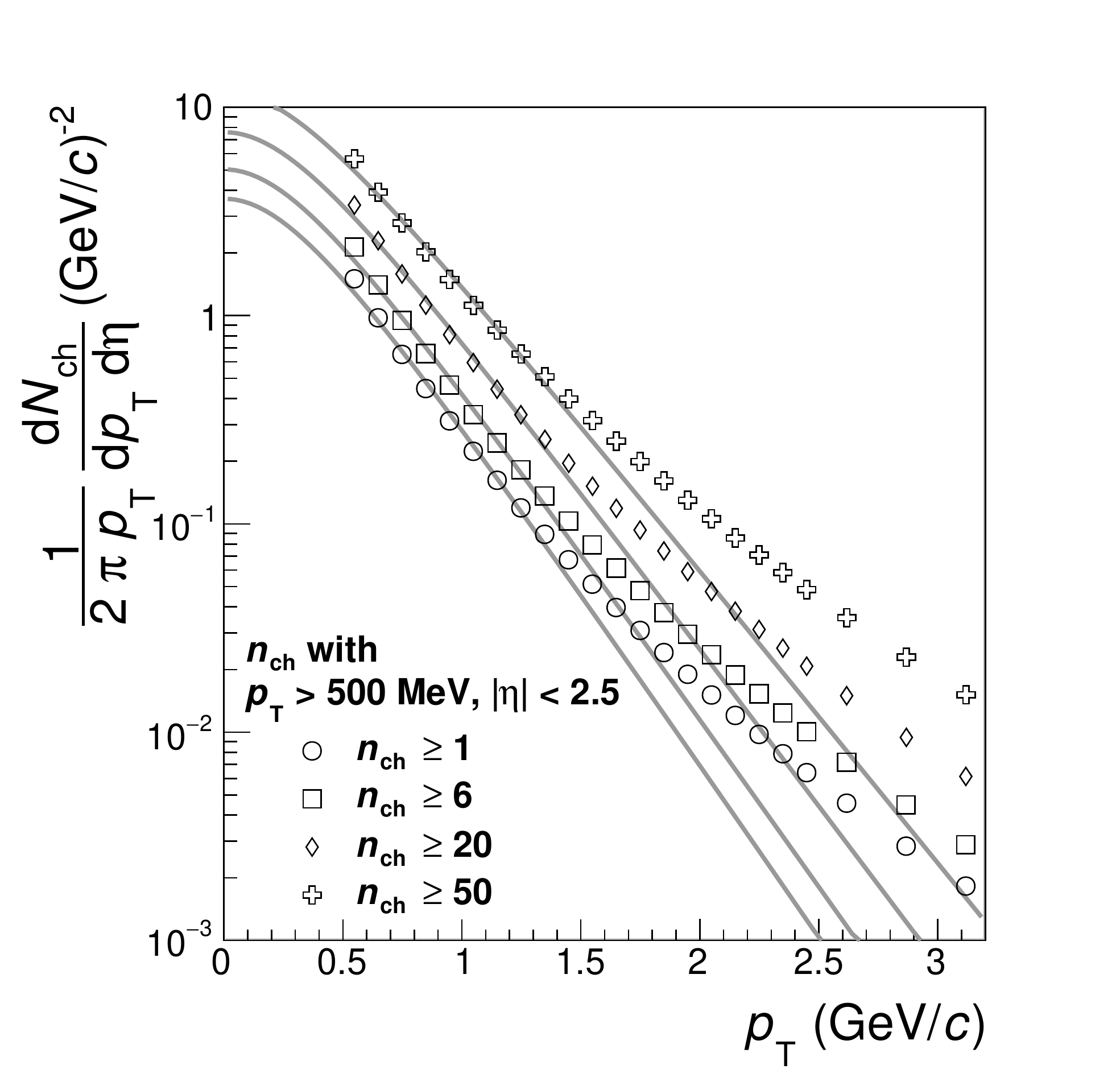}
\caption{Fit of Eq.~\ref{eq:meson_spectrum} (resulting from Gaussian fluctuations of $\lambda = \sqrt{\kappa}$ where $\kappa$ is the string tension) to charged-hadron spectra for different multiplicity classes measured by ATLAS \cite{Aad:2016xww}. }
\label{fig:fits}
\end{figure}

The ATLAS collaboration has obtained charged hadron transverse momentum spectra in pp collisions at a center-of-mass energy 8~TeV in different charged-hadron multiplicity classes \cite{Aad:2016xww}, reaching high average multiplicities of $\D N_\mathrm{ch}/\D\eta \approx 30$ corresponding to more than 4--5 times the multiplicity in minimum bias pp collisions at this collision energy \cite{Adam:2015gka}. The charged-hadron transverse momentum spectra start at $p_{\perp} \approx \unit[0.55]{GeV}/c$. In order to determine the average string tension for each of the four multiplicity classes we fit the function of Eq.~\ref{eq:meson_spectrum} in the range $0.5 < p_\perp < \unit[1.4]{GeV}/c$ to the meson spectra.  In Fig.~\ref{fig:fits} we show the fit of Eq.~\ref{eq:meson_spectrum} to the ATLAS charged-hadron transverse momentum spectra \cite{Aad:2016xww}.
One sees that the pion transverse momentum spectrum gives a decent description of the data in the range $0.5 \lesssim p_\perp \lesssim \unit[1.5]{GeV}/c$.
 
\begin{figure}
\centering
\includegraphics[width=0.95\linewidth]{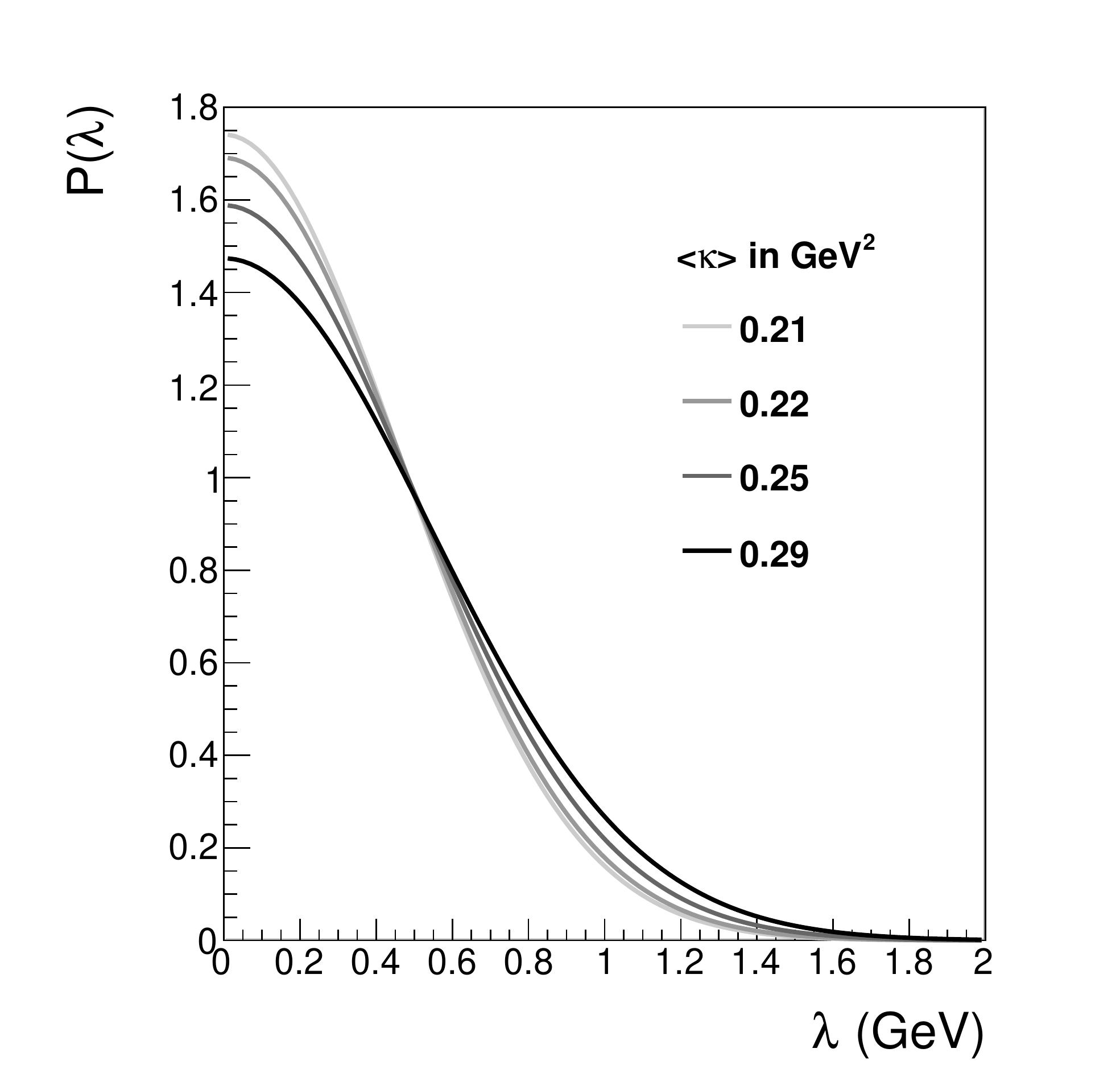}
\caption{Gaussian distributions $P(\lambda)$ as a function of $\lambda$, i.e., the square root of the string tension $\kappa$, for strings created in pp collisions for the four multiplicity classes of Tab.~\ref{tab:mult}.}
\label{fig:fluct_gaus}
\end{figure}

To estimate the total average charged-hadron multiplicity of each class we extrapolate the measured spectrum to $p_\perp = 0$ using a Tsallis function \cite{Tsallis:1987eu,Bhattacharyya:2017hdc}. In Fig.~\ref{fig:fluct_gaus} we show the corresponding distributions of $\lambda$, the square root of the string tension, for the four multiplicity classes. The curve for the lowest multiplicity is the most narrow. For higher multiplicities the widths of the Gaussian fluctuations increase. A higher average string tension corresponds to a smaller average diameter of the flux tube. The Schwinger mechanism takes into account pair creation by tunneling in a time-independent electric field, which is proportional to the average string tension \cite{Casher:1978wy}. The field itself is determined by the charges at the end of the string and the transverse area of the flux tube which can fluctuate \cite{Abramovskii:1997pc}.

We  can determine from the value of the mean string tension parameter the quark spectra Eq.~\ref{eq:quark_spectrum} for light and also for strange quarks when we have a value of the strange quark mass $m_s$. In order to agree with  \Py\ simulations for minimum bias collisions with $\langle \kappa \rangle = 0.21\,\mathrm{GeV}^2$ we fit the strange quark mass to the ratio of produced strange to light quarks $s \bar{s} / u \bar{u}= 0.217 $ which is used by \Py. We find a value 
\begin{equation}
  m_\mathrm{s} = \unit[0.687]{GeV}.
  \label{eq:quark_mass-strange}
\end{equation}

This value is higher than the kaon mass indicating a strong binding in the kaon system. With the light and strange quark constituent masses and the values of mean $\kappa$ for each multiplicity we can calculate $s \bar{s} / u \bar{u}$ ratios for the other multiplicity classes as shown in Table~\ref{tab:mult}. 

\begin{table}[h]
\begin{tabular}{c c c}
$(\D N_\mathrm{ch} / \D\eta)_{\eta=0}$ \quad & \quad $\langle \kappa \rangle$ in $\mathrm{GeV}^2$ \quad &  $s \bar s / u \bar u$\\
\hline
7.92 & 0.21 & 0.217  \\
11.87& 0.22 & 0.230  \\
18.8 & 0.25 & 0.257  \\
31.7 & 0.29 & 0.292 
\end{tabular}
\caption{Charged-hadron multiplicities at midrapidity in pp collisions at 8~TeV, mean string tension in $\mathrm{GeV}^2$, and the resulting ratio $s \bar s / u \bar u $.}
\label{tab:mult}
\end{table}

Fluctuations of the string tension to larger values lead to an increase of multiplicity of all hadronic species, but the relative increase is larger for strange hadrons than for pions. Besides, in high multiplicity pp collisions many flux tubes are produced. The high density of strings leads to a squeezing of the tubes, i.e., their transverse dimension is reduced and the mean string tension increases, because the area shrinks. 

Flux tubes repel each other. This can be easily understood \cite{Pirner:2014zka}. A $3$-$\bar{3}$ string has a string tension $\kappa_3=(2\pi\alpha_{\mathbb{R}}^\prime)^{-1}\approx 1 \mathrm{GeV}/\mathrm{fm}$ \cite{Casher:1978wy}, where $\alpha_{\mathbb{R}}^\prime\approx 0.9\,\mathrm{GeV}^{-2}$ is the slope of the Reggeon ($\bar qq$) trajectory. The energy per unit lengths of two such isolated strings is twice larger, about $2\,\mathrm{GeV}/\mathrm{fm}$. If, however, the two $3$-$\bar{3}$ strings merge and form a color octet-octet string, its string tension is related to the slope of the Pomeron trajectory $\kappa_8=(2\pi\alpha_{\mathbb{P}}^\prime)^{-1}\approx 4\,\mathrm{GeV}/\mathrm{fm}$. Therefore a merging of the strings leads to a significant increase of the energy stored in the string, i.e. to string repulsion. Moreover, the popular value $\alpha_{\mathbb{P}}^\prime=0.25 \, \mathrm{GeV}^{-2}$ used for this estimate, is a result of unitarity saturation in elastic $pp$ scattering. A more realistic Pomeron slope $\alpha_{\mathrm{P}}^\prime\approx 0.1\,\mathrm{GeV}^{-2}$ extracted from HERA data for diffractive electroproduction of vector mesons, would lead to a much stronger repulsion.

\begin{figure}
\centering
\includegraphics[width=0.95\linewidth]{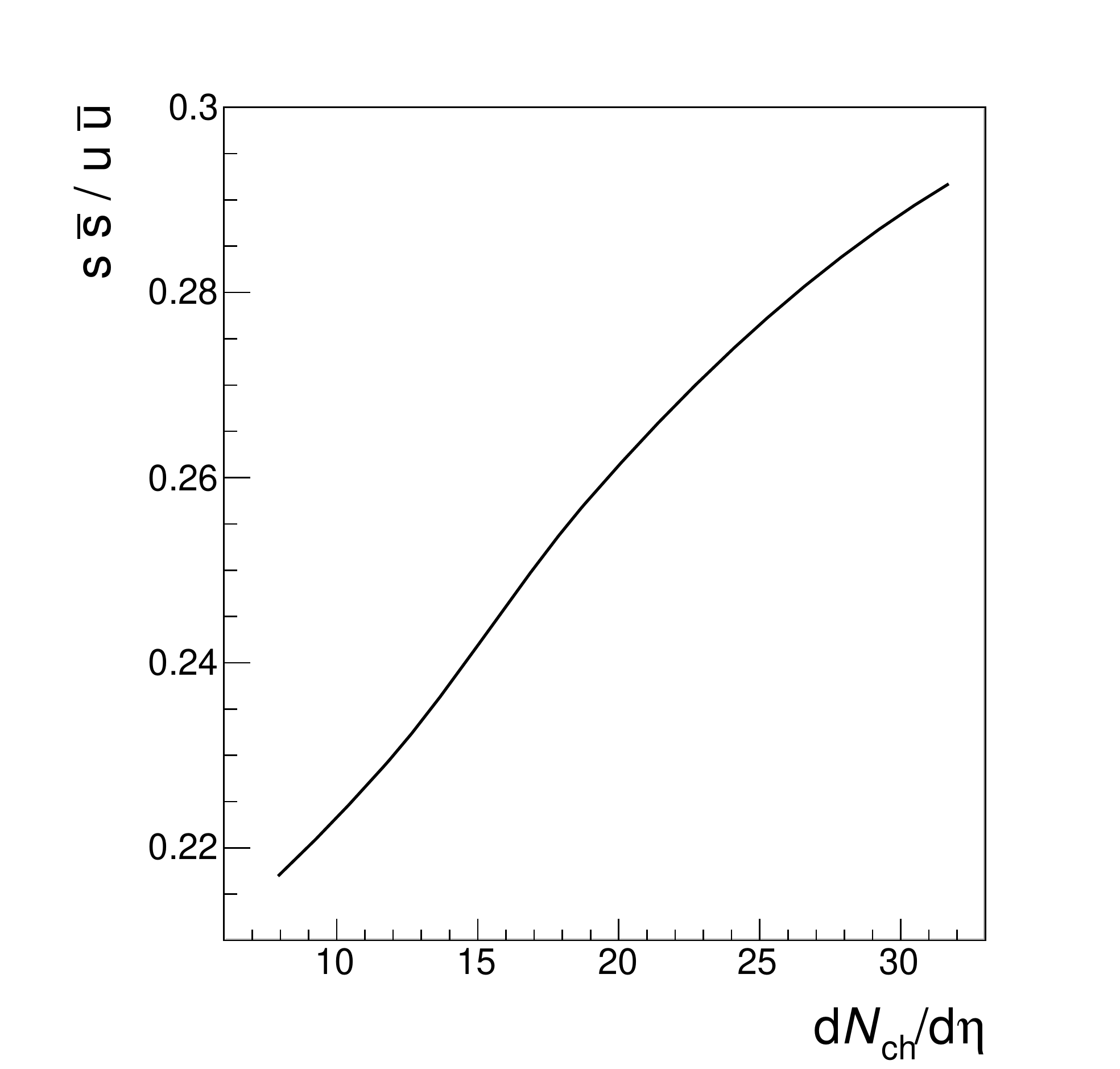}
\caption{The ratio of strange to light quark production, $s \bar{s}/ u \bar{u} $, produced in pp collisions vs charged-particle multiplicity.}
\label{fig:stoud}
\end{figure}

An increase of the central multiplicity $\D N_\mathrm{ch}/\D \eta$ of produced charged hadrons by a factor of four corresponds to an increase of the average string tension by about $30\%$. This increase generates more quarks, but strange quarks are more affected than light quarks. In Fig.~\ref{fig:stoud} we show the relevant increase of strange $ s\bar{s} $ relative to $ u\bar{u}$ quark production obtained from the integration of Eq.~\ref{eq:quark_spectrum} over transverse momentum, inserting as quark masses the strange quark mass and the light (u,d) quark masses of Eqs.~\ref{eq:quark_masses} and ~\ref{eq:quark_mass-strange}. 
Of course, the result depends on the constituent quark masses for the two flavors. 

It is important to realize that averaging the string tension changes the form of the mass dependence  in the exponential.  The conventional Schwinger formula gives a function with a quadratic mass dependence which is converted into a linear mass dependence by the string fluctuations.  This change and the increase of the average string tension leads a reasonable increase of the ratio of strange to light quarks from the default value $ s\bar{s} /u\bar{u} =0.217$ which is called \texttt{StringFlav:probStoUD} in \Py\ to  $ s\bar{s} /u\bar{u} =0.3 $ for high multiplicity pp-collisions.

\section{Strange hadron enhancement from \Py\ simulations}
\begin{figure}
\centering
\includegraphics[width=0.95\linewidth]{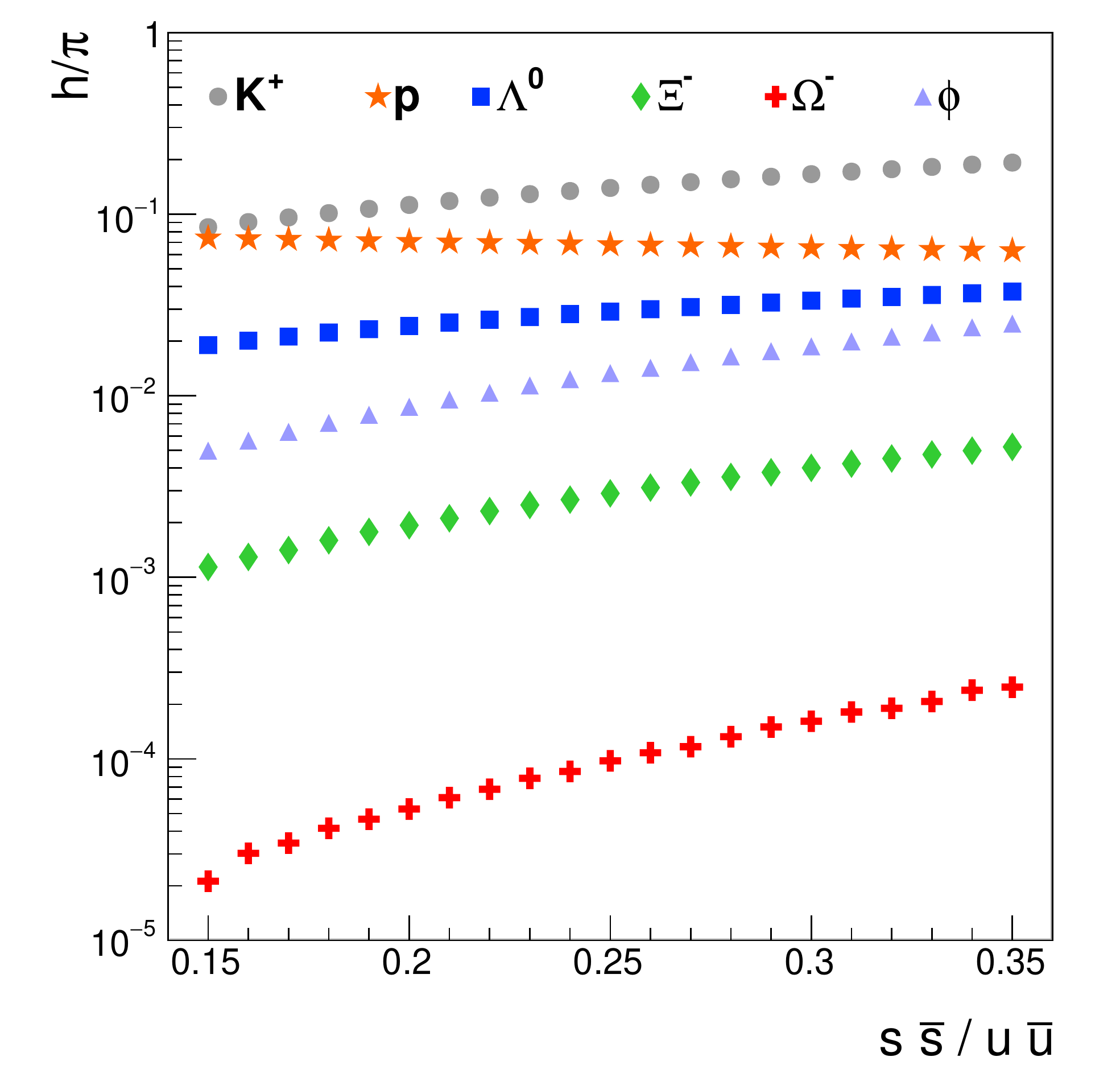}
\includegraphics[width=0.95\linewidth]{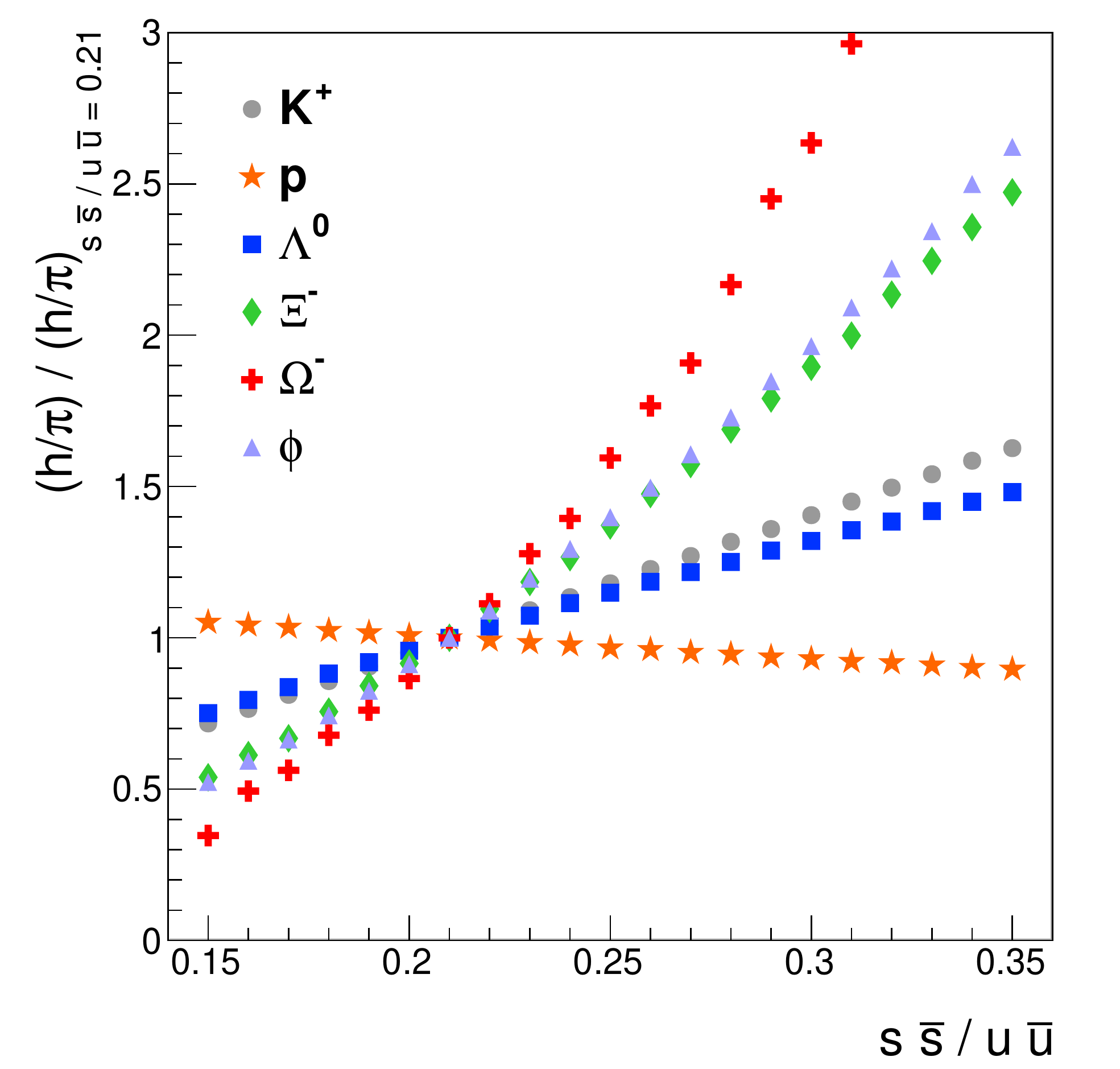}
\caption{Top: Hadron to $\pi^+$ ratio in the fragmentation of one string as simulated by \Py~8.235 as a function of the parameter
$s \bar s / u\bar u $, i.e., of the \Py\ parameter \texttt{StringFlav:probStoUD}. Bottom: Double ratio of the hadron to $\pi^+$ ratio $\mathrm{h}/\pi$ normalized to the value obtained for $s \bar s / u\bar u = 0.21$. The double ratio shows that the relative increase of the hadron yield increases with the strange quark content of the hadron.}
\label{fig:pythia_ratios_vs_stoud}
\end{figure}

Having obtained ratios of strange to light quark production from the Schwinger model with fluctuating string tension we can simulate the complicated hadronization process using the conventional string fragmentation. For this purpose we repeatedly simulate with \Py~8.235 the decay of one string spanned by a $u \bar u$ quark pair where each quark has an energy of 0.5~TeV. The yields of different hadrons normalized to the yields for $\pi^+$ are shown in Fig.~\ref{fig:pythia_ratios_vs_stoud} as a function of the strange-to-light quark ratio. One can see that the relative increase of the hadron to pion ratio becomes stronger with increasing strange content of the hadron. This is in qualitative agreement with the observations of the ALICE experiment \cite{ALICE:2017jyt}. From the double ratio (cf.\ Fig.~\ref{fig:pythia_ratios_vs_stoud} bottom panel) one sees nicely that the increase of each hadron species depends on the number of strange and antistrange particles it has. 
$\Omega$ baryons with three strange quarks are enhanced most, then come the $\Xi$ baryons and and $\phi$ mesons. Finally $K^+$ mesons and $\Lambda$ baryons  are enhanced in the same way. One sees that  the proton production rate  decouples almost from the strange particle production. We especially point out the behavior of the $\phi$ mesons. \Py, according to the OZI rule \cite{Okubo:1963fa,Zweig:1964jf,Iizuka:1966fk}, assumes that the s-quark and anti s-quark of the $\phi$ come from independent production processes. In the thermal model of meson production $\phi$ mesons are not especially suppressed, since their net strangeness is zero, and the thermal model only considers the masses of hadrons and a chemical potential for strange hadrons.

\begin{figure}
\centering
\includegraphics[width=0.95\linewidth]{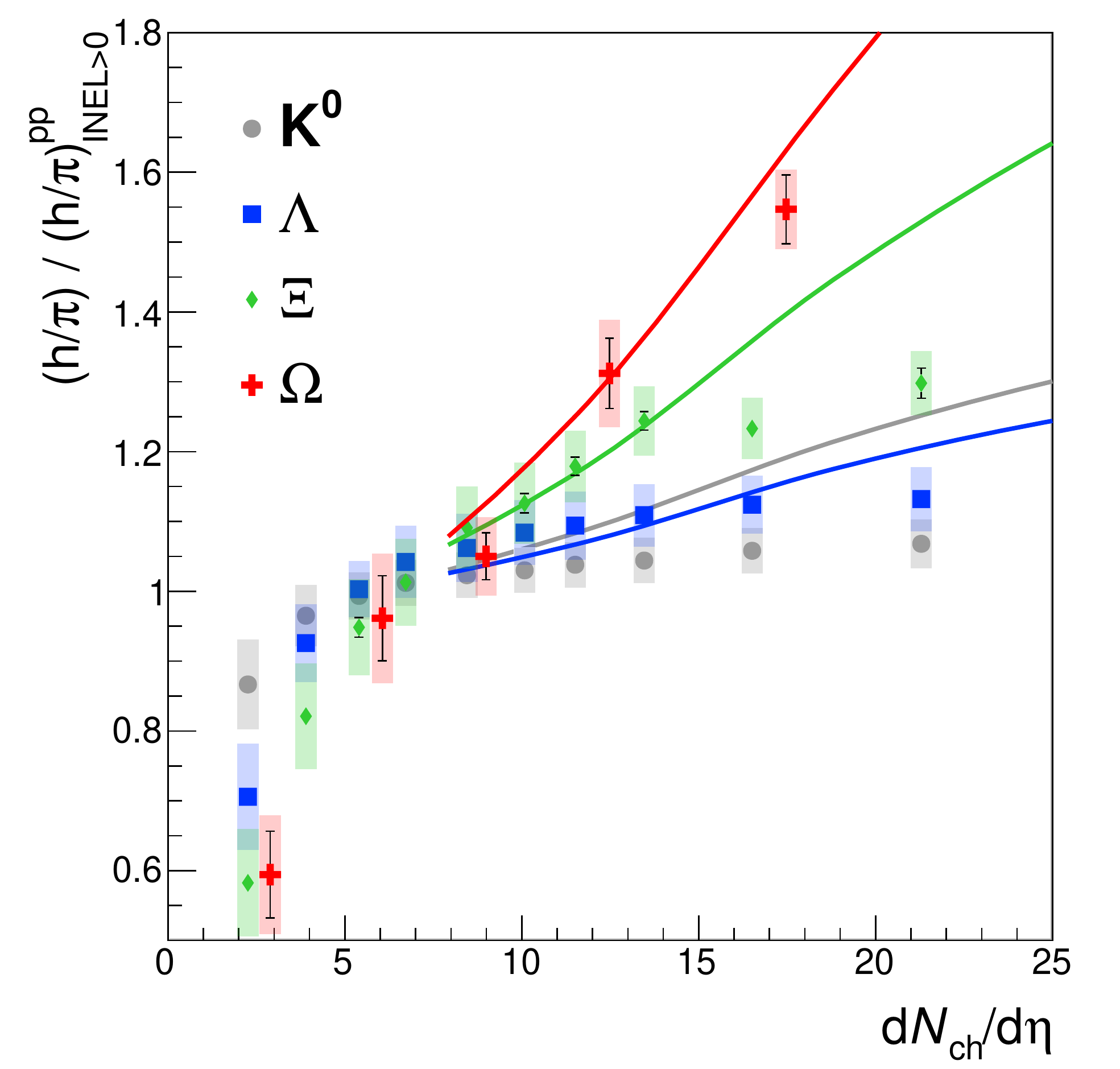}
\caption{The double ratio of strange hadron to light hadron production at high multiplicity to minimum bias multiplicity for K, $\Lambda$, $\Xi$, and $\Omega$ produced in pp collisions. The data are from the ALICE experiment (pp collisions at $\sqrt{s} = 7\,\mathrm{TeV}$, \cite{ALICE:2017jyt}). From top to bottom are the curves for the $\Omega$, $\Xi$, K, and $\Lambda$ double ratios.}
\label{fig:hadronratio}
\end{figure}

The dependence of the double ratios on $s\bar s / u \bar u$ production affects its dependence on the multiplicity $dN_\mathrm{ch}/d\eta$ (cf.~Fig.~\ref{fig:hadronratio}). The strongest theoretical increase  is obtained for the $\Omega$, in agreement with the data. We slightly overestimate the increases of the other strange hadrons with multiplicity.
 
\begin{figure}
\centering
\includegraphics[width=0.95\linewidth]{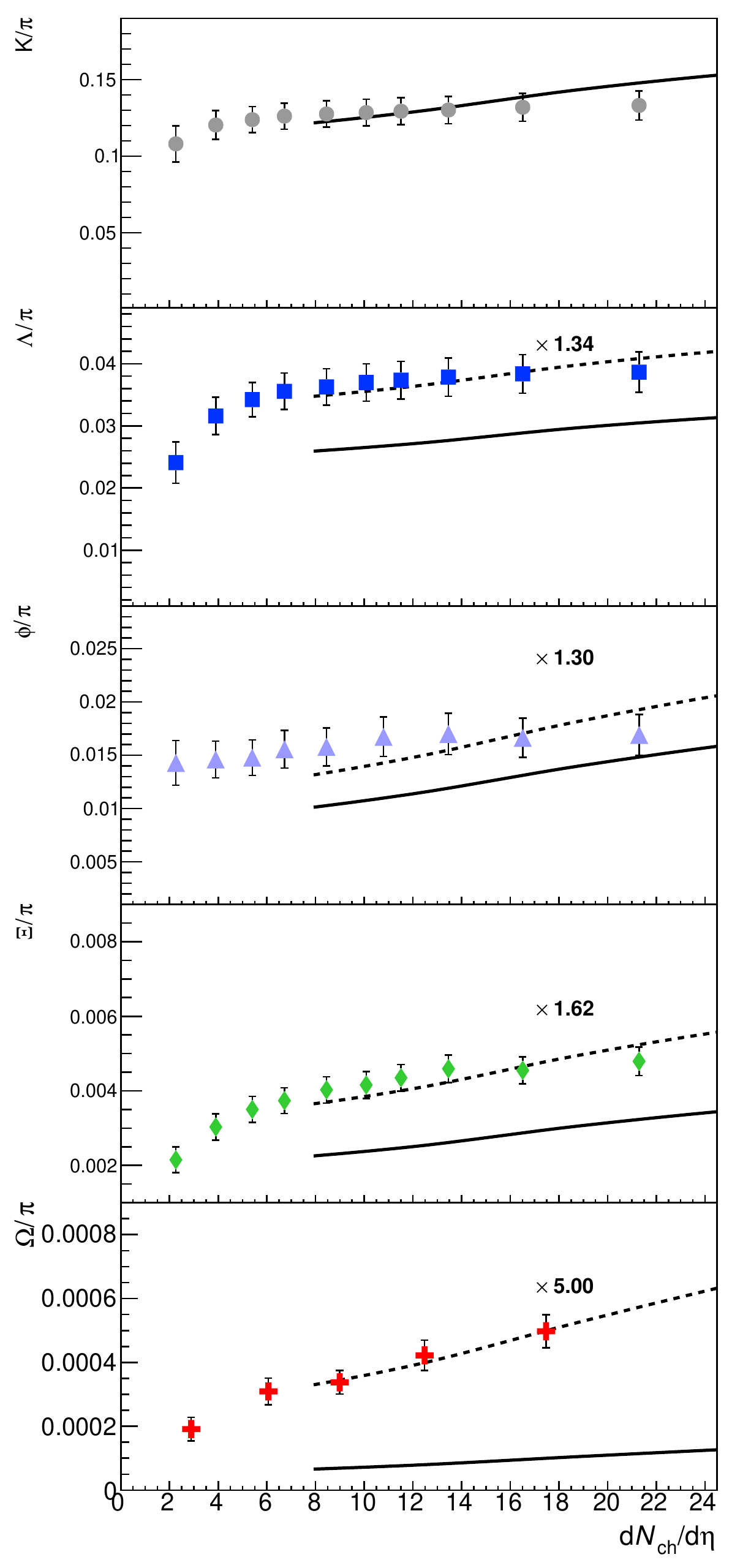}
\caption{The ratio of strange hadron to light hadron production ($\mathrm{K}^+/\pi^+$, $\Lambda^0/\pi^+$, $\phi/\pi^+$, $\Xi^-/\pi^+$, and $\Omega^-/\pi^+$) in pp collisions for different charged-hadron multiplicities $\D N_\mathrm{ch}/\D\eta$. The data measured in pp collisions at $\sqrt{s} = 7\,\mathrm{TeV}$ are from the ALICE experiment \cite{ALICE:2017jyt,Acharya:2018orn}. The predictions of the model presented in this paper (solid lines) were rescaled by the indicated factors to obtain the dashed lines.}
\label{fig:particle_ratios}
\end{figure}

The experimental data on absolute ratios confirm the increase of strange hadrons with multiplicity. One sees in Fig.~\ref{fig:particle_ratios}  that the theoretical  production ratios for different strange particles obtained from the simplified \Py\ simulation  follow the data over four orders of magnitude, which is a great achievement of the \Py\ simulation. But \Py\ falls short of the absolute data already for minimum bias collisions. In general \Py\ has a tendency to overshoot proton production and underestimate strange baryons. Note we use the standard diquark parameters from \Py\ 8.2 and  do not change the diquark production. We were not sure how to parametrize the necessary $uu$, $ud$, $dd$, $su$, $sd$, and  $ss$ diquark masses and how to implement them consistently in the code. A naive fit of diquark masses to the default values for baryon production used in \Py\ gave too large diquark masses. 

In this respect, we emphasize that our main focus here is on the effect of the fluctuating string tension, which enhances the strangeness production. In order to single out this effect from other model assumptions, we have considered a simplified model of  string fragmentation.
For this reason one should not expect a precise description of data, as we see in Fig.~\ref{fig:particle_ratios}.  There are several lines of further improvements on the phenomenology of strangeness and baryon production:\\
(i) In high multiplicity events the contribution of multistring final state configurations is enhanced. This increases the production rate of two or three strange quarks from decays of different strings. \\
(ii) The \Py\ mechanism of baryon production via diquark-antidiquark creation  is oversimplified,  if one assumes a similar Schwinger mechanism as in quark-antiquark creation. An alternative stepwise mechanism was proposed in \cite{Casher:1978wy}, which does not contain the diquark mass as a parameter. Instead of a $\bar qq$ pair, which completely screens the color field of the decaying string, a fluctuation can produce a $\bar qq$ of "wrong" color which is followed by the production of another $\bar qq$ pair, that finally completes the screening of the  color field of the string.\\
(iii) The principal difference between the baryon and meson topological structures is present in the form of a string junction \cite{Rossi:1977cy}, which is the genuine carrier of the baryon number. Thus, production of a baryon-antibaryon pair is unavoidably preceded by creation of a string junction-antijunction. This can be preformed by pure gluon mechanism, proposed in \cite{Kopeliovich:1988qm}. This mechanism, missed in \Py\, might be essential and might solve the problems in the description of baryon production. 

\section{Discussion and Conclusions}
We have presented a simple and schematic model for strangeness enhancement in high multiplicity pp collisions. We modified the result of the Schwinger model by letting the string tension $\kappa \equiv \lambda^2$ entering the Schwinger formula fluctuate according to a Gaussian distribution for $\lambda$. Using as experimental input the observed charged-particle transverse momentum spectra of hadrons at various multiplicities we have determined the effective average string tension entering the Gaussian string fluctuations at each multiplicity. Higher multiplicities correspond to higher average string tension. This modified Schwinger model gives definite $s \bar{s} / u \bar{u}$ ratios which then serve as inputs into simulations of the string fragmentation code \Py~8.2. The resulting hadronic rates reproduce qualitatively the increase of hadron production rates measured in the ALICE experiment \cite{ALICE:2017jyt}.

Quark production is always suppressed exponentially. In the Schwinger model the exponent depends quadratically on the quark mass. In our modified model with a fluctuating string tension the suppression is only linear in the exponent, thus enhancing the production of heavier quarks relative to light ones. We note that a scenario with merging flux tubes \cite{Brogueira:2006nz,Bierlich:2014xba,Nayak:2018xip} leading to a higher string tension can also enhance strangeness. It appears to us, however, that phenomenological fits \cite{Fischer:2016zzs} favor the linear mass dependence we generate in our model. Finally, the quite successful thermal models always include a linear suppression of heavier flavor states in the exponent. In the thermal picture it remains a challenge to find good theoretical reasons for quick apparent thermalization in spite of the small volume created in pp collisions. An attempt in this direction has been made in Ref.\cite{Berges:2017hne} where for extremely short times such a suppression is obtained in the bosonized Schwinger model. Recently, a preliminary investigation \cite{PirnerFloerchinger} using a time dependent electric field created by color sources with different transversal structure give rates which after integration over all times yield a linear mass suppression in the exponent. In the future, the study of pp and pA collisions with short reaction times extended over small volumes make it an experimental and theoretical task to understand the phenomenon of early ``thermalization''. Another main phenomenological question remains the necessity to simulate the baryon rates adequately.
Notice that high multiplicities are usually reached due to multiple interactions, producing multi-string final state configurations. In terms of Glauber-like models this correspond to large number of unitary cut Pomerons in the case of pp collisions \cite{Abramovsky:1973fm}, or to large number of participants in the case of nuclear collisions \cite{Miller:2007ri}. Coherence in such collisions leads to the effects of shadowing and saturation which are stronger for light than for heavy flavors. This is why, e.g., the production rate of $J/\psi$ rises steeply with light hadron multiplicity \cite{Kopeliovich:2013yfa,Kopeliovich:2019phc}. Some enhancement remains for strangeness production, although not as pronounced as for charm. For the sake of clarity we ignored here these additional mechanisms of strangeness enhancement, but concentrated on the single string dynamics.

\appendix*

\section{Production of mesons with wave function}
In this short appendix we present a more elaborate calculation on the production rate of mesons based on the rate for quark and antiquark production and a Gaussian meson wave function $\phi(q_\perp)$ in transverse momentum space. The  calculation confirms the dependence estimated in the simplified model of Sec.II, as we will show:

\begin{figure}
\centering
\includegraphics[width=0.95\linewidth]{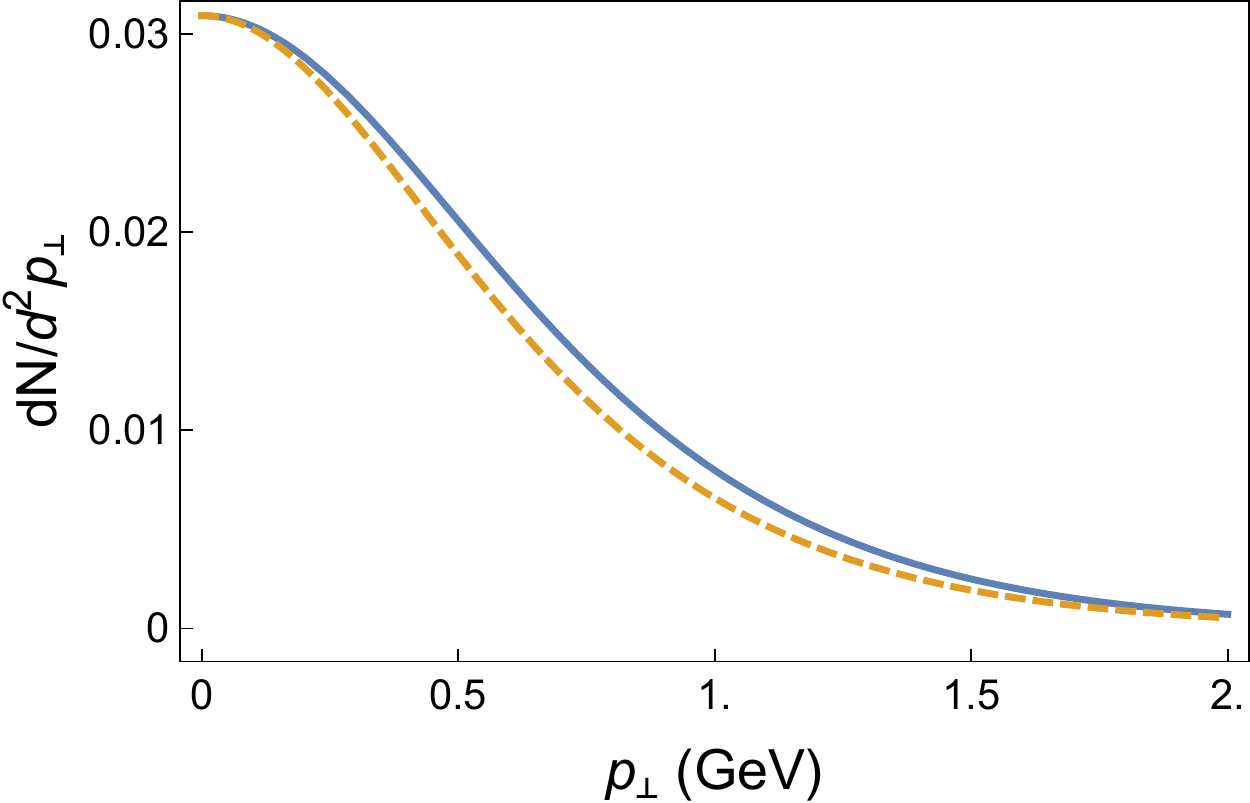}
\caption{The differential meson cross section is shown as a function of transverse momentum $p_\perp$ [GeV] for the calculation with wave function (blue) and for the simple Gaussian model (orange dashed).}
\label{fig:appendix}
\end{figure}

\begin{equation}
|\phi(q_\perp)|^2=\frac {r^2}{\pi}e^{-r^2 q_\perp^2}.
\end{equation}
For later numerical comparison the mean momentum in the wave function is $\langle q_\perp^2 \rangle =\frac{1}{r^2}= \unit[0.25]{GeV^2}$.
The meson production rate then has the following form in this picture:
\begin{align}
\frac{dN}{d^2p_\perp} = &\int e^{-\beta \sqrt{p_{\perp,1}^2 + m^2}} e^{-\beta \sqrt{p_{\perp,2}^2+m^2}} \frac{r^2}{\pi}e^{-r^2 (p_{\perp,1}-p_{\perp,2})^2} \nonumber \\ 
& \delta(\vec{p}_{\perp,1}+\vec{p}_{\perp,2}-\vec{p}_\perp) d^2p_{\perp,1} d^2p_{\perp,2}
\end{align}
with
\begin{equation}
\beta=\sqrt{\frac{2 \pi}{\langle \kappa \rangle}}.
\end{equation}
We use the trick to convert the square root dependence of the quark production rate into a Gaussian dependence with the help of the integral representation:
\begin{equation}
e^{-\beta \sqrt{p_\perp^2+m^2}}=\frac{2}{\sqrt{\pi}}\int_0^\infty \mathrm d\mu \, e^{-\mu^2-\frac{\beta^2}{4 \mu^2} (p_\perp^2+m^2)}.
\end{equation}
Then
\begin{align}
\frac{dN}{d^2p_\perp} = & \frac{r^2}{\pi} \int \mathrm d\mu_1 \mathrm d\mu_2 e^{-\mu_1^2-\frac{\beta^2}{4 \mu_1^2}(p_{\perp,1}^2+m^2)} e^{-\mu_2^2-\frac{\beta^2}{4 \mu_2^2}(p_{\perp,2}^2+m^2)}\nonumber \\
& e^{-r^2 (p_{\perp,1}-p_{\perp,2})^2} \delta(\vec{p}_{\perp,1}+\vec{p}_{\perp,2}-\vec{p}_\perp) d^2p_{\perp,1} d^2p_{\perp,2}.
\end{align}
We define  new variables:
\begin{align}
\vec{p_\perp} &= \vec p_{\perp,1} + \vec p_{\perp,2} \\
\vec{q_\perp} &= \frac{\mu_2^2 \vec p_{\perp,1} -\mu_1^2 \vec p_{\perp,2}}{\mu_1^2+\mu_2^2}
\end{align}
and use for the maximum of the integral with $\mu_1=\mu_2$ the approximate relation:
\begin{equation}
\vec{q} \approx \frac{\vec{p}_{\perp,1} - \vec{p}_{\perp,2}}{2}.
\end{equation}
Further evaluation follows the steps given in Ref.\cite{Pirner:1997kc}. We end up with
\begin{equation}
\frac{dN}{d^2p_\perp} = \frac {2 \sqrt{2}}{\pi} \frac{\sqrt{(p_\perp/2)^2+m^2}}{4 \sqrt{(p_\perp/2)^2+m^2} + \frac{\beta}{r^2}} e^{-2 \beta \sqrt{(p_\perp/2)^2 + m^2}}.
\label{eq:meson_inv_yield_wave_fct}
\end{equation}
In Fig.~\ref{fig:appendix} we show a comparison of the meson spectra calculated with meson wave function 
Eq.~\ref{eq:meson_inv_yield_wave_fct} with the simple Gaussian function in Eq.~\ref{eq:meson_spectrum} normalized appropriately. We see that for the relevant momentum region the two calculations agree quite nicely. 

\begin{acknowledgments}
This work is part of and supported by the DFG Collaborative Research Centre ``SFB 1225 (ISOQUANT)''. B.K. is partially supported by grants ANID - Chile FONDECYT 1170319, by ANID PIA/APOYO AFB180002, and by USM internal project PI\_LI\_19\_13.
\end{acknowledgments}



\bibliography{stringreferences}






\end{document}